\documentclass{IEEEtran}

\usepackage{graphicx}
\usepackage{tabularx}
\usepackage{graphicx}
\usepackage[british]{babel}
\usepackage{eurosym}
\usepackage{pbox}
\usepackage{url}
\usepackage{xspace}
\makeatletter
\g@addto@macro{\UrlBreaks}{\UrlOrds}
\usepackage{booktabs}
\renewcommand\IEEEkeywordsname{Keywords}
\usepackage[font=small,labelsep=space]{caption}
\usepackage[noadjust]{cite}
\def\citepunct{,\penalty\citepunctpenalty\hskip.13emplus.1emminus.1em\relax}
\def\citedash{\hbox{--}\penalty\citepunctpenalty}

\begin{document}

\title{The Perfect Storm: \\The Privacy Paradox and the Internet-of-Things}

\author{\IEEEauthorblockN{Meredydd Williams, Jason R. C. Nurse and Sadie Creese\\}
\IEEEauthorblockA{Department of Computer Science\\
University of Oxford\\
Oxford, UK\\
\{\textit{firstname.lastname}\}@cs.ox.ac.uk}}

\maketitle

\begin{abstract}
Privacy is a concept found throughout human history and opinion polls suggest that the public value this principle. However, while many individuals claim to care about privacy, they are often perceived to express behaviour to the contrary. This phenomenon is known as the Privacy Paradox and its existence has been validated through numerous psychological, economic and computer science studies. Several contributory factors have been suggested including user interface design, risk salience, social norms and default configurations. We posit that the further proliferation of the Internet-of-Things (IoT) will aggravate many of these factors, posing even greater risks to individuals' privacy. This paper explores the evolution of both the paradox and the IoT, discusses how privacy risk might alter over the coming years, and suggests further research required to address a reasonable balance. We believe both technological and socio-technical measures are necessary to ensure privacy is protected in a world of ubiquitous technology.
\end{abstract} 


\section{Introduction}
\label{sec:one}

Privacy has been an important concept throughout human history, with many great civilisations and philosophers considering the subject. The Code of Hammurabi protected the Ancient Babylonian home against intrusion by others \cite{Dash2004}, while Socrates distinguished between the `outer' and `inner' self \cite{Konvitz1966}. Warren and Brandeis placed privacy in the modern democratic consciousness, reacting to perceived excesses in photojournalism by defining a `right to be let alone' \cite{Warren1890}. Privacy is now enshrined as a legal and human right in many nations across the world, particularly through Article 17 of the International Covenant on Civil and Political Rights. With privacy considered essential to both democracy \cite{Lefort1988} and natural human development \cite{Berscheid1977}, it is of little surprise that many claim to value this liberty.

Numerous opinion polls and surveys suggest that individuals care about privacy. In 2015, the University of Pennsylvania found that 84\% of participants want to control disclosure to marketers, with 91\% disagreeing that data should be traded for customer discounts \cite{Turow2015}. A 2013 Pew Research Center poll similarly found that 86\% of participants reported taking steps to remain private online, whether by cleaning cookies or encrypting emails \cite{Rainie2013}. While these studies suggest that individuals value their privacy, there is much evidence to the contrary. Carrascal \textit{et al.} found participants were willing to sell their web browsing history for only \euro 7 \cite{Carrascal2013}, while Beresford and colleagues discovered that individuals neglect privacy concerns while making purchases \cite{Beresford2012}. Researchers found that 74\% of US respondents had location-based services enabled, exchanging sensitive information for convenience \cite{Zickuhr2012}. This presents the `Privacy Paradox' \cite{Brown2013}, where individuals claim to value their privacy but appear to not act accordingly. Previous work has suggested that a number of factors, including user interface design \cite{Masiello2009}, risk salience \cite{Hughes-Roberts2014} and privacy settings \cite{Debatin2009}, can exacerbate this disparity between claim and action.

The Internet-of-Things (IoT) promises to be the digital revolution of the twenty-first century \cite{vasseur2010}. It has the potential to connect together a vast number of ubiquitous components, enmeshing itself within our everyday lives. It promises to offer a wealth of opportunities for productivity and convenience and is predicted to generate trillions of dollars for the global economy \cite{manyika2015}. While the revolutionary appeal of the IoT is clear, its development is likely to be in tension with privacy. Wearable fitness devices have already suggested the pregnancy of their owners \cite{Marcus2016}, while connected TVs can eavesdrop on background conversations \cite{BBCNews2015a}. Small, ubiquitous products enable pervasive data collection at a scale far greater than previously possible. Constrained gadgets communicate remotely with other heterogeneous appliances, with owners having little understanding of their novel products.

We posit that the further development of the IoT will exacerbate the Privacy Paradox, causing a number of privacy risks for consumers. Those factors which these new devices aggravate, such as the salience of risk, mental models and default settings, are precisely those which currently contribute to this phenomenon. In this paper we look to explore how the Internet-of-Things will fundamentally differ from conventional computing technologies and how this will impact the Privacy Paradox. From this, we suggest research avenues, both technological and socio-technical, which we believe will promote a reasonable balance between privacy and functionality. We are, to our knowledge, the first work to consider the paradox in this novel context and hope to elucidate the privacy risks which these new technologies bring.

The remainder of our paper is structured as follows. Section \ref{sec:two} surveys the Privacy Paradox in detail, considering opinion polls which suggest concern, evidence to the contrary, and existing literature. Section \ref{sec:three} then discusses those factors considered contributory to the paradox, including lack of user awareness, interface design and privacy policy complexity. Section \ref{sec:four} explores the IoT before examining the significant novelties of these products compared to conventional computing devices. Section \ref{sec:five} discusses the intersection of the topics and the technological and socio-technical research we believe necessary for the future. Finally, we conclude the paper in Section \ref{sec:six} and reflect on what implications the Internet-of-Things might have for privacy.

\section{The Privacy Paradox}
\label{sec:two}

As previously discussed, a large number of opinion polls and surveys have shown that individuals claim to value privacy. In 2013, Pew Research Center found that 86\% of surveyed US citizens reported taking steps to remain private online, with actions ranging from ``clearing cookies to encrypting their email'' \cite{Rainie2013}. A University of Pennsylvania poll concluded that 84\% of US participants ``want to have control over what marketers can learn about'', also finding that 91\% disagree that data collection is a fair trade for consumer discounts \cite{Turow2015}.

Research undertaken in the aftermath of the surveillance revelations has shown privacy concerns on both sides of the Atlantic. Researchers discovered 87\% of US respondents had heard of the scandal with 34\% of those altering their behaviour \cite{Rainie2015}, while a customer concern survey found 92\% reported worrying about their privacy online  \cite{TRUSTe2015}. Although the general public clearly claim to care about their privacy, they are often found to act contrary to their reports.

Carrascal and colleagues used a reverse second price auction to analyse the values placed on personally identifiable information (PII) \cite{Carrascal2013}. They discovered participants were willing to sell their browsing history for only \euro 7, contrasting with the oft-claimed importance of privacy. PII is rapidly becoming an outdated concept, with aggregations of publicly-available data now often of greater sensitivity than personal details. Beresford \textit{et al.} conducted a similar study, instructing individuals to buy from one of two competing online stores, with the first requiring greater disclosure than the second \cite{Beresford2012}. In spite of this, almost all participants chose the first store when it was \euro 1 cheaper, and proportions were equal when the prices were identical. In addition to low valuations, individuals often act promiscuously with their personal data. 

The popularity of social networking websites such as Facebook and Twitter have led many to share excessive amounts of information online \cite{Symantec2010}. While location-based services are convenient for navigation, one poll found almost one in five participants ``checked-in'' at physical locations, enabling social media observers to track their exact whereabouts \cite{Zickuhr2012}. In a 2016 survey, although two-thirds of respondents reported to want greater privacy protection, only 16\% used protective plug-ins and less than one-in-ten encrypted their emails \cite{HideMyAss2016}.  Although the general public might claim to value their privacy, they are found to act in a paradoxical manner.

This disparity between what individuals claim about privacy and how they actually act is known as the Privacy Paradox \cite{Brown2013}. Although the situation might appear illogical on initial inspection, the existence of this phenomena has been suggested through a number of studies. Barnes analysed the social networking habits of US teenagers and concluded that ``adults are concerned about invasions of privacy, while teenagers freely give up personal information'', attributing this disparity to adolescents' lack of awareness \cite{Barnes2006}. Acquisti and Gross surveyed Facebook users in analysing the impact of privacy concerns on observed behaviour \cite{Acquisti2006}. Validating the paradox, they found even those with significant concerns joined the network and shared large amounts of data. 

Norberg \textit{et al.} questioned participants on their willingness to disclose data before requesting the same information through market researchers twelve weeks later \cite{Norberg2007}. They found that regardless of the type of information, including PII and financial data, respondents disclosed a far greater quantity than they initially claimed. Acquisti found that individuals act differently to what had been traditionally considered rational, concluding that users focus on the short-term gratification of a service without considering the long-term risks \cite{Acquisti2004a}. The following year, Acquisti and Grossklags surveyed a number of students on their privacy attitudes, finding that while almost 90\% claimed to be concerned, their usage of protective technologies was ``consistently low''. \cite{Acquisti2005}. These studies repeatedly indicate a disparity between the claimed value of privacy and the actions which individuals undertake to protect it. From this, a number of factors have been suggested which contribute to the paradox.

\section{Contributory Factors}
\label{sec:three} 

The prevalence of the Privacy Paradox has been frequently investigated, but of greater utility is understanding which factors contribute to this phenomenon. By investigating what leads to this disparity between claim and action, we can look to better-protect individuals' privacy. Through surveying existing literature we identified five classes of factors which compound the paradox: \textit{education and experience}, \textit{usability and design}, \textit{privacy risk salience}, \textit{social norms}, and \textit{policies and configurations}. These categories are neither intended to be exhaustive nor mutually exclusive: some factors could be placed in multiple classes while others were omitted due to their immutability. For example, demographics are found to have an influence, with women seen to be more privacy-conscious than men \cite{Sheehan2002}. However, that certain groups perceive privacy differently is a product of largely immutable physiological and sociological factors which cannot be easily altered. We continue by discussing these classes of contributory factors and considering issues in the context of the Internet-of-Things.

\subsection{Education and Experience}

Education has been shown to affect individuals' perceptions of privacy. O'Neil analysed an online survey and found that those with doctoral degrees possessed the greatest level of privacy concern, successively followed by vocational degrees, professional degrees, college attendance and high school \cite{ONeil2001}. Williams and Nurse saw that those with the highest levels of education revealed the fewest elements of optional data \cite{Williams2016}. In their study of demographic data disclosure they went on to find that those educated in cybersecurity matters were even more reluctant to reveal their information. Lewis and colleagues discovered that those with more online experience are likely to have stronger privacy configurations \cite{Lewis2008}, suggesting digital literacy has an effect. This might be for a number of reasons, including the importance of self-efficacy and that those most acquainted with computing devices are likely to feel less intimidated by technology. While desktops and laptops might be familiar to a large section of society, the proliferation of novel IoT devices could pose a greater challenge as heterogeneous products flood the market.

\subsection{Usability and Design}

Adams and Sasse explained how individuals do not try to act insecurely, but poor usability is an impediment to correct behaviour \cite{Adams1999}. When users misjudge system functionality they often place their privacy and security at risk, whether by misconfiguring application settings or divulging information accidentally \cite{Strater2008}. Individuals might possess mature mental models of how they expect a computing device to function; when these assumptions are misplaced then issues can arise \cite{Prettyman2015}. Therefore, those well-trained in operating conventional computers might misjudge the functionality of novel IoT technologies such as smart appliances or wearable devices.

While individuals might develop mental models which correctly align with user interfaces, understanding systems-level interactions will be more challenging. Traditionally data was collected by a single device and stored locally, or only shared under the explicit consent of the user. However, our world is becoming increasingly interconnected as information is collated and aggregated in vast quantities. Individuals might disclose data in one context without considering the consequences of further propagation and dissemination. For example, while students proudly share photos of themselves across social media, they can later regret these decisions when their images are viewed by potential employers. With IoT nodes surreptitiously recording and sharing input from their surroundings, ordinary users might have little knowledge of how widely their data has spread.

Unfortunately, many online platforms are specifically designed to maximise information disclosure. Ulbricht evaluated Facebook design through the lens of institutional economics, observing a conflict of interest as the portal wishes to collect as much data as possible \cite{Ulbricht2012}. This appears also true for IoT technologies, with user information providing an abundant resource for monetisation. Jensen \textit{et al.} found that interfaces which display trust marks reduce privacy concern \cite{Jensen2005}, and social networking sites receive large amounts of personal information partly due to their attractive design. The novelty and functionality of nascent IoT devices might similarly distract consumers from the quantities of data they are disclosing.

\subsection{Privacy Risk Salience}

The salience of privacy risk is also an important factor affecting user behaviour. Spiekermann \textit{et al.} noted concerns might differ between the online and offline world, with even private individuals forgetting their reservations in digital environments \cite{Spiekermann2001}. While citizens might intuitively regard a closed door as protecting one's privacy, they have difficulty interpreting equivalent actions in the online world \cite{Creese2009}. Even if individuals do become acquainted with the importance of passwords and cyber security, privacy risks can be disguised by the novelty and functionality of new devices. Numerous studies have suggested the importance of salience, with Tsai \textit{et al.} finding that privacy indicators on search engines can encourage individuals to alter their behaviour \cite{Tsai2009}. Adjerid \textit{et al.} found that even a delay of 15 seconds between a privacy notice and a decision can result in less-private actions being taken \cite{Adjerid2014}. If the salience of privacy risk can be obscured by familiar computing devices, ubiquitous IoT technologies will only exacerbate this problem.

\subsection{Social Norms}

Social norms play an important role in defining what individuals consider normal and acceptable. While consumers of the 1980s stored personal data in their own homes, now many think little of sharing their lives across social media and the cloud. Broad changes in attitudes can lead to herding effects, where individuals feel compelled to align with the actions of their contemporaries \cite{Devenow1996}. This goes some way to explain why privacy-protective messaging services such as Signal fail to gain market share, as  users are unwilling to invest in niche products not used by their friends. Metcalfe's Law \cite{Metcalfe1995} states that the value of a network is proportional to the square of its connected users, and therefore privacy apps face a challenge in gaining initial support. Internet users move ``like a swarm of killer bees'', adapting their behaviours to match that of those around them \cite{Solove2007}. In such a context, even when individuals wish to act privately, their behaviour gravitates to what they consider ``normal''. Social norms often differ between cultures, and privacy perceptions have been shown to vary across the world \cite{Alashoor2015a}. Daehnhardt \textit{et al.} conducted a study of Twitter settings, finding that citizens from Japan were more private than those from Brazil or Spain \cite{Daehnhardt2015}. This was due to individuals from `Multi-Active' societies being considered more likely to project their opinions than those from `Reactive' cultures.

Norms also differ across age groups, with research suggesting teenagers disregard their privacy \cite{Barnes2006}. However, children have been shown to still value this principle in other contexts, such within the family \cite{Cranor2014}. The elderly often encounter the Privacy Paradox, possessing strong concerns despite facing numerous technological obstacles \cite{Paine2007, Smith2014}. Irrespective of culture or age, as IoT devices integrate themselves into society, the acceptability of ubiquitous data collection is likely to increase. Utz and Kr{\"a}mer found that social network settings were adjusted as users undertook impression management, often reducing protections to promote themselves more effectively \cite{Utz2009}. As Rose aptly stated, ``society is changing, norms are changing, confidentiality is being replaced by openness'' \cite{Rose2011}, and this trend looks to continue with the Internet-of-Things.

\subsection{Policies and Configurations}

While privacy policies should increase transparency and reduce disparities between expectation and reality, the opposite is often true. In the Jensen \textit{et al.} study, they found that concern was reduced by the mere presence of privacy policies, even if they were not read \cite{Jensen2005}. Policies are frequently written in such an obfuscated fashion that even users who care about their privacy might have little recognition of the data they are disclosing \cite{Bashir2015}. Acceptance is gradually becoming more implicit, moving from click-wrap licensing through checking a box, to browse-wrap licensing by reading a webpage \cite{Lesk2015}. Individuals eager to use a service are likely to ignore privacy statements, sacrificing sensitive information before they understand their loss. In a similar fashion, those enthusiastic to use novel IoT devices might bypass documentation and become bound to conditions of which they have little awareness.

Default configurations might not respect privacy and rely on the inertia of users to support data collection. While social networks might provide extensive privacy settings for customisation, Mackay found that individuals tend not to deviate from default configurations \cite{Mackay1991}. Even when attempts at alteration are made, controls are frequently too complex for ordinary users to make meaningful progress \cite{Johnson2012}. Compounded, this creates ecosystems where many users possess insufficient privacy protection, even though opportunities are demonstrably available \cite{Govani2005}. If current user interfaces and configurations are troublesome for the general public, this will only be exacerbated through the proliferation of IoT devices.

\section{The Internet-of-Things}
\label{sec:four}

The Internet-of-Things (IoT) has the potential to be a truly revolutionary technology \cite{vasseur2010}, connecting together vast numbers of devices and blurring the boundaries between the virtual and the physical. Miorandi and colleagues described it as a ``global network interconnecting smart objects'', ``the set of supporting technologies necessary'' and the market opportunities leveraged from these developments \cite{miorandi2012}. In essence, the IoT is the interconnection of large numbers of ubiquitous devices which collect and process data. Although our current Internet is vast, its scale will pale in comparison to these exciting new networks \cite{gartner2014}.

The IoT has a long history of development through a number of related fields, including computer networking, telecommunications and wireless sensor networks \cite{vasseur2010}. However, the IoT looks to differ from many conventional devices, particularly servers, desktops and laptops. From a reflection on the existing literature, we have identified five classes in which new developments might depart from existing technologies: \textit{usability and configuration}, \textit{ubiquity and physicality}, \textit{resource constraints}, \textit{unfamiliarity and heterogeneity}, and \textit{market forces and incentives}. Again, these categories are neither intended to be exhaustive nor mutually exclusive; for example, product unfamiliarity and poor usability could both contribute to the formation of inaccurate mental models. However, these classes act as a useful scaffold to explore how the IoT might alter the technological landscape and what impact this might have on privacy.

\subsection{Usability and Configuration}

Due to small form factors and low unit prices, IoT user interfaces might not be as rich or expressive as found on modern desktop computers \cite{Rowland2015}. In 2013 the UK government commissioned a usability study on a series of IoT heating devices, finding that none of the five market leaders offered sufficient user interfaces \cite{Department2013}. Issues included the ``complex setup of schedules'' and ``difficulty identifying system state'', both of which could contribute to the inadvertent disclosure of sensitive data. While existing technologies are familiar to a large proportion of individuals, those using IoT devices might not possess the accurate mental models \cite{Kranz2010} required for their security and privacy.

Even those able to navigate novel interfaces might not understand the summation of Internet-of-Things interactions. Two decades ago data was stored on local hard-drives, whereas current individuals are more prepared to save their information remotely. However, as IoT nodes communicate autonomously and forward aggregations to distant networks, ordinary users might not be conscious of where their data is going. Cloud services already cause concerns for corporations \cite{Zhou2010}, who might not know where their information is stored, and these trends will only continue. While a device in isolation might align with existing mental models, the composability of IoT products poses a novel conceptual challenge.

Just as free social networks need to collect user information, IoT technologies are primarily designed to facilitate interaction and automation. Individuals seldom deviate from default settings \cite{Mackay1991}, and this inertia might increase in IoT environments. While apathy already reduces the likelihood of amendment, IoT reconfiguration might be constrained by limited interfaces or require advanced technical knowledge \cite{University2015}.

\subsection{Ubiquity and Physicality}

Although the scale of the current Internet is vast, it will appear minute in comparison to IoT infrastructure \cite{gartner2014}. While some devices might already possess online capabilities, such as CCTV cameras, the number of such technologies will increase on an exponential scale over the coming decades. This produces several challenges, including uniqueness and addressability, with billions of machines connected to the same infrastructure. Fortunately, the development of IPv6 provides a suitably large address space and makes Network Address Translation (NAT) largely redundant.

The scale of data collection will also be unprecedented \cite{Herold2015}. While modern mobile phones collect metrics through their accelerometers and gyroscopes, pervasive IoT technologies could document our lives as never before \cite{Gemmell2014}. Although the online and offline worlds are increasingly enmeshed, we can usually isolate ourselves from the Internet by undertaking our tasks in an analogue fashion. However, as sensors record their surroundings and actuators alter their environments, the distinction between the physical and the virtual becomes increasingly unclear \cite{Bruner2014}. 

\subsection{Resource Constraints}

While modern machines appear virtually unrestricted in their computing power, many IoT products operate with very constrained resources. The origin of this issue is twofold: firstly, IoT gadgets require a portable power source for remote environments; and secondly, small form factors do not allow for large batteries \cite{Blaauw2014}. To maximise their lifespan, these devices constrain their processing power, limiting the complexity of possible calculations \cite{Ma2011}. These restrictions dictate that resource-intensive communication protocols are infeasible, posing a challenge for secure transmissions which rely on cryptography \cite{Roman2011a}. If IoT devices communicate using weak or immature protocols then data confidentiality is placed at risk.

\subsection{Unfamiliarity and Heterogeneity}

Whereas individuals are accustomed to desktop and laptop computers, only a small proportion so far are familiar with IoT products \cite{Accenture2015}. With these new technologies offering significant benefits to convenience and productivity, consumers might be myopic to the potential risks of the platform. Although we use numerous operating systems and hardware from many vendors, the range of IoT devices is unprecedented, turning ``heterogeneous networks'' into ``super-heterogeneous networks'' \cite{Grindrod2015}. The IoT is an incredibly nebulous term, including but not limited to home automation (e.g. Nest, Hive), wearable devices (e.g. Fitbit, Apple Watch) and industrial control. This extreme heterogeneity proves problematic for standardisation, leading to several piecemeal approaches which do not seamlessly interact \cite{Bandyopadhyay2011}. Although the IPSO Alliance looks to unify vendors from across technology and communications, further work is required to enable widespread interoperability.

\subsection{Market Forces and Incentives}

Since products which are launched early establish strong commercial positions, appearance and functionality is often prioritised over other considerations. The implication for the IoT is that the market will be flooded with attractive and feature-rich devices, with privacy and security only considered as an afterthought \cite{Coates2014}. Internet-of-Things products are frequently sold at low unit prices, and these will only decrease further as competition expands. Efforts to reduce manufacturing costs have implications for what might be viewed as the non-essential functionality of the device. For example, a small smartwatch might not have mechanisms to set a password or PIN code. It is unlikely inexpensive products will possess strong tamper-resistance, posing risks especially when nodes are stationed in remote locations. Few of us currently own IoT devices \cite{Accenture2015} and therefore their ubiquity and necessity does not appear inevitable. However, herding effects suggest \cite{Devenow1996} that once the IoT is firmly established individuals will begin considering these technologies as ``normal'' and abstention might be viewed as antiquated.  

\section{Discussion and Future Research}
\label{sec:five}

The Internet-of-Things will disrupt our established notions of technology, both offering numerous opportunities and promoting a series of risks. Developments of the next decade will transform many factors which current socio-digital interactions rely on. While the Privacy Paradox presently leads to challenging disparities between claim and action, we posit that the aforementioned contributory factors will be aggravated by the IoT. This exacerbation will occur for three key reasons, \textit{The Interfaces}, \textit{The Data} and \textit{The Market}, as discussed below.

\subsection{The Interfaces}

Firstly, novel, heterogeneous and often-constrained user interfaces might lead to further privacy risks. Usability is critical for the correct operation of a device, with mistakes and misconceptions contributing to costly errors. In an age where few bother to peruse user manuals, technology must carefully align with mental models to offer intuitive interaction. However, while the IoT market is rapidly expanding, few of the general population have yet purchased these promising devices \cite{Accenture2015}. Therefore, these technologies will remain unfamiliar for some time and during this period existing mental models might be misaligned, what Karl Smith terms ``cognition clash'' \cite{Smith2016}. In this manner, even if individuals care about their privacy they might have little idea of how to protect their data. 

The heterogeneity of devices in the nebulous IoT will also contribute to this issue, as users struggle to familiarise themselves with a miscellany of differing interfaces. Currently consumers might select operating systems from several vendors or hardware from a dozen manufacturers, but they are soon familiar with these technologies. However, with many IoT products possessing small or non-existent screens, individuals might struggle to become acquainted. The best efforts of privacy-conscious individuals might be in vain if there is little understanding of how to correctly use a device.

The affordances and constraints of an interface subtly direct a user to undertake certain actions and refrain from others \cite{Norman1988}. With both the core functionality and underlying business models of IoT products relying on data collection, devices might be designed to encourage disclosure. Default settings, seldom adjusted \cite{Mackay1991}, will likely enable the capture of data which can be monetised by vendors and third parties. While user apathy contributes to reconfiguration inertia, IoT privacy settings might be too obfuscated to offer a feasible alternative. Even if individuals wish to protect their data, they might require specialist technical knowledge to reconfigure their products, if this is even possible.

Risk is a salient thought in the physical world, with societies developing norms and customs to minimise the dangers they face. Virtual threats are less tangible as individuals feel insulated behind their computer screens \cite{Jackson2005a}; a factor which has contributed to the growth of cybercrime. Risk salience has been found critically important in encouraging privacy-conscious actions online \cite{Tsai2009}, and this is likely true in novel IoT environments. Citizens understand that locking a door protects their privacy and  gradually accept that strong password selection achieves a similar goal in the virtual space. However, in environments as nebulous, novel and heterogeneous as the IoT, ordinary individuals will have little chance understanding the risks they face, irrespective of their privacy concerns.

\subsection{The Data}

Secondly, ubiquitous device presence and unprecedented levels of data collection will expand the reach of technological surveillance. This is partially due to the vast scale predicted for the IoT, with billions of devices embedded into every aspect of our society \cite{gartner2014}. The US government have already confirmed that IoT devices could be the target of surveillance efforts, presenting the strategic importance of these ubiquitous nodes \cite{Ackerman2016}. Although our current networking infrastructure is large and widespread, it does not pervade our lives in the manner suggested by the nascent Internet-of-Things.

Another contributory factor to this surveillance is the ubiquitous presence of devices, continuing the gradual shift from the server room to the bedroom to the pocket. Although smartphones and tablets succeed through their extended mobility, it is wearable devices that begin to blur the physical and virtual worlds. When nodes monitor bodily functions and sensors track their owners, privacy becomes antiquated regardless of calls for protection.

Although data collection can be conducted covertly, we usually have some degree of awareness. Whether through unnoticed CCTV operation signs or voluminous privacy policies, we generally have some means of detecting surveillance. However, when ubiquitous devices pervade our physical world we cannot be certain when our actions are being monitored. Outcry was seen in response to Google Glass functionality, with restaurants refusing patrons who might be covertly recording each other \cite{Page2013}. Even discounting state surveillance, device vendors could quickly infer one's daily patterns, diet and social interactions \cite{Nurse2014}. With inaccurate mental models and constrained user interfaces obstructing device configuration, consumers might be oblivious to when products are actually active. Even those who profess to value their privacy can do little when they are oblivious to pervasive monitoring.

Of considerable concern is the lack of consent required by ubiquitous devices. Although clickwrap and browsewrap licenses are ignored by most online visitors, they still present opportunities for users to review the terms to which they are bound. This assumes that the party navigating to a website is the one whose data is collected, but this might not be true in IoT environments. While those who purchase devices can examine privacy policies, those without IoT products can be monitored without their consent. In this manner, those most fervently in opposition to surveillance might still have their privacy violated by the actions of device owners.

\subsection{The Market}

Thirdly, market forces and misaligned incentives will contribute to the proliferation of cheap devices with minimal privacy protection. As these products become more lucrative, competition will expand and prices will decrease. While smart appliances might have larger price points, wearable gadgets and sensor nodes will possess tighter profit margins. With the market driven by novelty, appearance and functionality, manufacturers will have little incentive to offer strong security or privacy \cite{Coates2014}. To the contrary, user data can be monetised and sold to third parties, suggesting privacy is in tension with the business models of many IoT vendors. With security features requiring valuable time and money to implement, manufacturers may rationally invest those resources in enhancing functionality, especially when consumers are seen to use services in spite of their privacy concerns \cite{Acquisti2006}. These incentives suggest a proliferation of insecure data-collecting devices, with privacy-conscious individuals offered little alternative.

While international privacy standards could offer protection against the IoT, their guidelines are often unrealistic. ISO/IEC29100 requests companies ``try and understand the expectations and preferences'' of their users \cite{InternationalOrganizationforStandardization2015}, a suggestion in direct tension with the Privacy Paradox. User expectations, intentions and actions often differ wildly, with this likely true in unfamiliar IoT environments. The framework also considers PII (Personally Identifiable Information) to not include anonymised data, despite the proven effectiveness of many de-anonymisation techniques \cite{Narayanan2008}. As with many international standards, companies are not compelled to comply with these guidelines. Therefore IoT vendors are likely to pursue their own financial interests and collect extensive quantities of data.

Software updates are an expensive fixed cost for a vendor, only supported by a profitable initial sale. Funding developers to undertake maintenance requires a healthy profit margin, which might not be feasible for smaller devices. At this point, manufacturers have little incentive to patch software vulnerabilities or improve the functionality of their shipped products. Again, this leads to large numbers of insecure devices which place the confidentiality of consumers' data at risk. With research suggesting that individuals trade privacy for a \euro 1 discount \cite{Carrascal2013}, secure alternatives might find difficulty gaining support, and therefore leave the market.

At the systems-level, the market is driving the increased composability of IoT devices. With intelligence moving from the centre to individual nodes, information will be collected and processed in unprecedented quantities. Whereas mundane data points might pose little risk in isolation, the aggregation of such metrics could reveal highly-sensitive details. Once information has been disclosed and propagated through IoT networks it might be impossible to amend or delete. Internet-of-Things products should not be considered in a vacuum; it is the multiplication of their individual functionalities which could place privacy in jeopardy.

As previously mentioned, appearance and functionality appears to drive the IoT market rather than privacy or security \cite{Coates2014}. Although adoption is currently tentative, once products reach a critical mass then herding effects and marketing will stimulate greater proliferation. The prospect of storing our personal information in an external data centre might have seemed radical two decades ago, but this is how many of us live our daily lives. Individuals share sensitive details with the world on social networking sites, even though similar actions in the 1980s would have perplexed contemporaries. Social norms might evolve more slowly than technology, but market pressures encourage the increased disclosure of personal information. While we might currently label smart-watches, smart-clothing or embedded health devices as unnecessary or invasive, these technologies could pervade society in the next decade. The disparity between privacy claim and action is great, partially fuelled by technological changes to established social practices. The Privacy Paradox will only grow as individuals clutch to traditional notions of privacy while living in a vastly interconnected world.

The interconnections between Internet-of-Things novelties and those factors contributory to the Privacy Paradox are summarised below in Figure \ref{fig:figure}.

\begin{figure}[ht]
  \centering
    \includegraphics[width=0.47\textwidth]{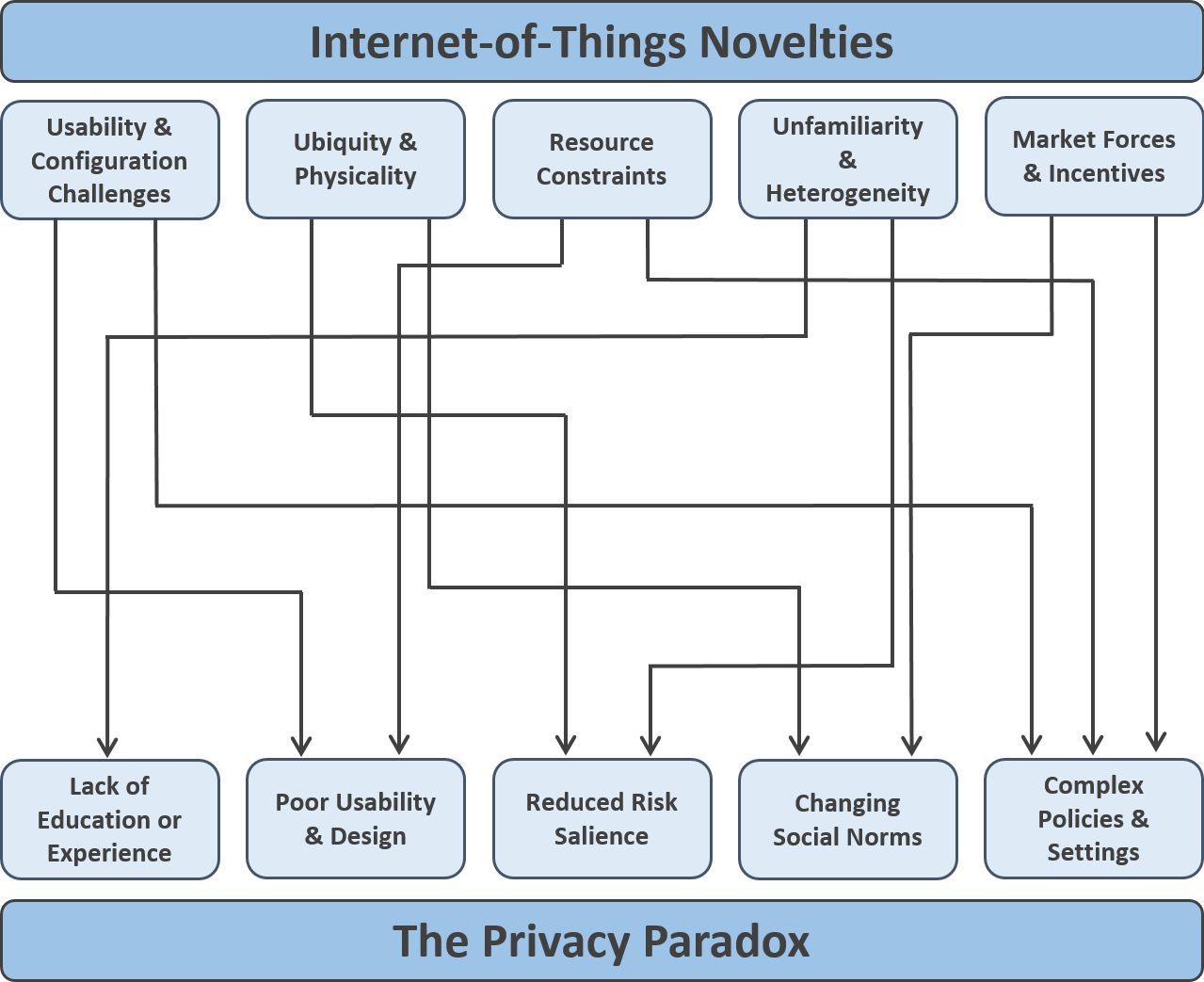}
    \caption{Internet-of-Things novelties in relation to the Privacy Paradox}
    \label{fig:figure}
\end{figure}

\vspace{-1em}

\subsection{Future Research}

We have identified a number of ways in which IoT developments might exacerbate the Privacy Paradox. If we wish to preserve privacy over the coming decades we must pursue research of both a technical and socio-technical nature. As technological investigations are often in tension with IoT vendor interests, software restrictions and Terms of Use could stifle several approaches. Several of the following studies could be constrained by proprietary architectures or privacy policies, and research should be considered on the basis of its likely practicality. However, with many devices offering rich APIs and the interactivity of products likely to increase, opportunities are set to expand in the future. Naturally, certain individuals might be willing to trade their privacy for convenience, with the principle being highly subjective and contextual. Our suggested research aims to enable individuals to make an informed choice rather than constrain their actions in a paternalistic manner.

Technologically, there are several endeavours which might help us better align privacy claims, intentions and actions. The research community could explore the development of enhanced user interfaces as a means of increasing risk salience. Prior research suggests that highlighting the concept of risk can encourage privacy-conscious behaviour \cite{Tsai2009}, and we could use a similar approach to alert IoT users when they face important decisions. Such techniques should not inhibit device functionality nor restrict individuals from performing actions, but simply inform them of the privacy risks they face. If we find that individuals using these enhanced interfaces disclose less sensitive data, then this suggests that marketplace IoT devices can be amended to better protect privacy.

In another approach to increase salience, risk exposure could be calculated and displayed based on disclosed information. Data points could be correlated and aggregated to infer unknown activities, using a similar approach to a data-reachability model \cite{Creese2012}. By highlighting the risk that individuals face through their IoT interactions, users could make informed choices on which features to enable. The product owner could define which pieces of information they wish to disclose, with the application traversing an inference tree to highlight what details they are actually revealing. Such an approach would look to reduce the disparity between individuals' perceptions of privacy risk and the consequences of their actions.

Default settings currently contribute to the Privacy Paradox, as discussed earlier. As a means of promoting privacy as a standard, configurations could be programmatically adjusted to reduce unnecessary data disclosure. While this might impair some device functionality, certain features could be permitted but considered opt-in rather than opt-out. With users seldom deviating from default configurations \cite{Mackay1991}, we could study whether the risk exposure of ordinary individuals can be reduced through these modifications.

We could explore the mocking of individual device readings by interacting with underlying APIs. Device functionality could be largely retained while faked values are provided in lieu of real data, protecting the privacy of concerned individuals. A similar technique was taken by the MockDroid tool, which interfaced with Android applications to reduce sensitive data leakage \cite{Beresford2011}. Although false metrics could reduce the usefulness of parts of a device, the addition of noise to individual readings could maintain an accurate average while concealing the true data. If users can appreciate product functionality while their personal information is masked, then this has implications for the development of IoT Privacy-Enhancing Technologies (PETs).

Of the most significant privacy concerns is that IoT devices can record data without the explicit knowledge of their owner. In an attempt to mitigate this risk, we could explore restricting surveillance to certain time-periods or geographic locations. This could be potentially achieved by modifying APIs, programmatically toggling configurations or jamming communications signals. User testing could complement this approach to ascertain whether individuals can receive some respite from incessant data collection. With few users expecting surreptitious surveillance, these modifications could better align privacy expectations with the reality.

Although technological approaches can support privacy, socio-technical research is also required for a comprehensive solution. Opaque privacy policies contribute to a disparity between individual expectations and reality. We could explore the development of concise and legible documents to accompany IoT products, and investigate whether this reduces the Privacy Paradox. If users understand how their personal data will actually be used, then perhaps they will be more cautious in disclosing their information. Individual privacy preferences could be encoded and compared against IoT products to highlight conflicts, in a similar manner to the P3P web tool \cite{LorrieCranor}. Web corporations have been previously compelled to simplify their privacy policies \cite{BBCNews2015} and such a mandate for smart devices should be considered by regulators.

As a means of counteracting market pressures for data collection, we could develop a metric for IoT privacy. This would consider the quantity of information extracted, the covertness of surveillance, the necessity of this data and how it is stored by the vendor. Those manufacturers who collect data for purely functional purposes might be rated highly, while those who sell aggregations to third parties would receive low scores. Establishing a competitive advantage for privacy, these metrics would recognise products which offer convenient functionality without demanding excessive amounts of personal data. This would be similar in principle to food health ratings, which shame restaurants with poor standards and incentivise vendors to improve their practices. 

Finally, the importance of device familiarity could be further explored by comparing the actions of novice and experienced IoT users. A longitudinal study could track disclosure levels over a period of time, investigating whether individuals alter their behaviour as they become more accustomed to a product. Device users might actually provide more personal information as they discover additional functionality or risk salience decreases. Such research would clarify our understanding of familiarity, of particular importance as the market is flooded with novel and heterogeneous products.

\section{Conclusions}
\label{sec:six}

In this paper, we have considered the Privacy Paradox: the disparity between what individuals claim about privacy and how they appear to act. We reviewed those factors which have been found contributory to this phenomenon, including user interface design, risk salience and social norms. We described the nascent field of the Internet-of-Things and considered how these novel technologies might differ from more-conventional computing devices. We posited that these developments will aggravate those same factors which contribute to the Privacy Paradox, further compounding a challenging situation. We claimed this exacerbation was for three key reasons: novel, heterogeneous and constrained user interfaces; ubiquitous device presence and vast data collection; and market forces and misaligned incentives. Finally, to further investigate the matter we suggested technological and socio-technical research, including enhancing user interfaces, altering default configurations and simplifying IoT privacy policies. 

It is our hope that such work would promote a reasonable balance between functionality and data protection, rather than accepting surveillance for convenience. Just as individuals share their information with greater freedom than two decades ago, future societies might view pervasive data collection as entirely normal. With market incentives encouraging aggregation and ignoring both privacy and security, regulation might be required to limit vendor violations. As the physical world becomes increasingly intertwined with the virtual, the tangibility of privacy risk could further decrease. While the Privacy Paradox leads individuals to disclose despite their concerns, when data collection is supported by billions of insecure devices there might not be any alternative.

\section*{Acknowledgment}
We wish to thank the UK EPSRC who have funded this research through a PhD studentship in Cyber Security.

\bibliographystyle{IEEEtran}
\bibliography{bib}

\end{document}